\documentclass[aps,prb,reprint,groupedaddress,showpacs]{revtex4-1}
\usepackage{graphicx}

\begin{document}


\title{Tunneling properties of vertical heterostructures of multilayer
  hexagonal boron nitride and graphene}

\author{Samantha Bruzzone} \affiliation{Dipartimento di Ingegneria
  dell'Informazione, \\Universit\`a di Pisa, Via G. Caruso 16, 56122
  Pisa, Italy.}
\author{Gianluca Fiori} \affiliation{Dipartimento di Ingegneria
  dell'Informazione, \\Universit\`a di Pisa, Via G. Caruso 16, 56122
  Pisa, Italy.}  \author{Giuseppe Iannaccone}
\affiliation{Dipartimento di Ingegneria dell'Informazione,
  \\Universit\`a di Pisa, Via G. Caruso 16, 56122 Pisa, Italy.}

\date{\today}

\begin{abstract}
We use first-principle density functional theory (DFT) to study the
transport properties of single and double barrier heterostructures
realized by stacking multilayer h-BN or BC$_{2}$N, and graphene films
between graphite leads. The heterostructures are lattice matched. The
considered single barrier systems consist of layers of up to five h-BN
or BC$_{2}$N monoatomic layers (Bernal stacking) between graphite
electrodes.  The transmission probability of an h-BN barrier exhibits
two unusual behaviors: it is very low also in a classically allowed
energy region, due to a crystal momentum mismatch between states in
graphite and in the dielectric layer, and it is only weakly dependent
on energy in the h-BN gap, because the imaginary part of the crystal
momentum of h-BN is almost independent of energy.  The double barrier
structures consist of h-BN films separated by up to three graphene
layers. We show that already five layers of h-BN strongly suppress the
transmission between graphite leads, and that resonant tunneling
cannot be observed because the energy dispersion relation cannot be
decoupled in a vertical and a transversal component.
\end{abstract}


\maketitle


Mobility of suspended graphene can be extremely high, as demonstrated
by experiments and theory~\cite{Bolotin,Du,Amad,Borysenko}.  However,
use of graphene in solid state devices typically requires deposition
or growth on a dielectric substrate, which can strongly suppress
mobility, due to electron coupling with dielectric-layer phonon modes,
as demonstrated in the case of SiO$_2$ or HfO$_2$, the most commonly
used dielectric materials~\cite{Morozov,Betti}.

Hexagonal boron nitride (h-BN) has been recently investigated as a
promising dielectric for graphene~\cite{Britnell2,Dean,Ponomarenko,Amet}, while boron nitride
domains has been suggested as a way to engineer graphene nanoribbon
transport properties~\cite{Lopez2012}.

h-BN has hexagonal geometry and a lattice constant closely
matching that of graphene~\cite{Meyer}, and the electronic interaction
of h-BN with a single graphite sheet~\cite{Dean,Decker} or with
bilayer graphene~\cite{Ramasubramaniam} is very weak.  As a result,
graphene deposited on an h-BN substrate maintains its electronic and
transport properties. 

From this perspective, understanding transport
along stacked graphene/h-BN structure can provide relevant information
regarding the leakage current in graphene based Field Effect
Transistors exploiting h-BN as gate dielectric or the performance 
of recently proposed vertical devices~\cite{Britnell,Guo}.

\begin{figure} [tbp]
\includegraphics[width=8cm]{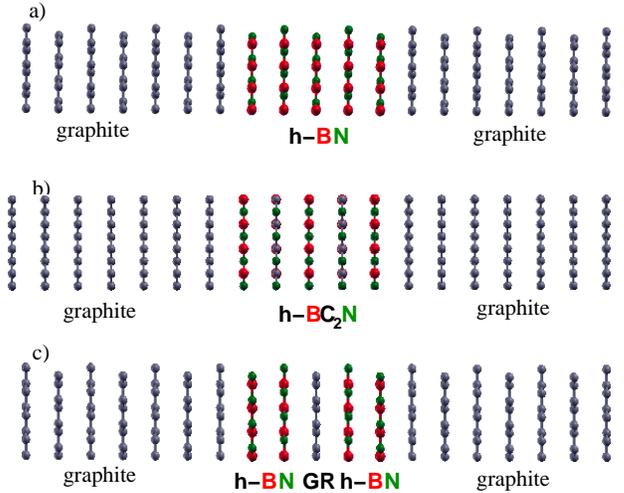}
\caption{Side view of the supercell used to represent the scattering
  region corresponding to a) graphite$|$5(BN)$|$graphite; b)
  graphite$|$5(BC$_2$N)$|$graphite; c)
  graphite$|$2(BN)$|$graphene$|$2(BN)$|$graphite.}
\label{geometry}
\end{figure}

In this work, with \textit{ab-initio} method, we calculate the quantum
transport in single and double barriers consisting of h-BN and BCN
layers between semi infinite graphite leads, using the formalism
introduced by Choi et al.~\cite{Choi}.  Barrier consists of 
up to five h-BN or h-BC$_{2}$N atomic layers arranged as shown in
Fig.~\ref{geometry}, according to a recent paper by
Ribeiro~\cite{Ribeiro}.  The double barrier is made of two or four
h-BN or h-BC$_{2}$N layers separated by up to three graphene layers.
The scattering region considered in the calculation includes at least
four atomic layers of graphite on each side of the barrier film, to
ensure that the charge density does not change at the interface
between the lead and the scattering region. The geometry of the total
structure in the scattering zone has been optimized.

Ab-initio calculations have been performed by means of Quantum
Espresso~\cite{QE}, using a plane wave basis set in the local density
approximation (LDA)~\cite{Perdew}.  A 35 Ry wave function cutoff has
been considered, the Brillouin zone has been sampled using a $30
\times 30 \times 30$ Monkhorst$-$Pack grid.  Atomic positions have
been relaxed using a conjugate gradient algorithm until all components
of all forces was smaller than 0.01 eV/\AA while the electronic
minimization has been performed with a tolerance of $10^{-7}$ eV.  It
has been shown~\cite{Giovannetti} that LDA provides a reliable
description of the geometry and the electronic structure in the
presence of weak interactions between h-BN layers or between h-BN and
graphene layers.  On the other hand, the LDA representation for the
exchange-correlation potential cannot correctly describe the excited
states in organic systems and leads to overestimation of
conductance. However, following other works on this
subject~\cite{Kuroda}, we expect that these effects have only minor
consequences, and thus our conclusions should not be affected by this
choice.  The transmission probabilities have been calculated with the
PWCOND~\cite{Smogunov} module of Quantum Espresso.  The transport
properties are studied in the framework of Landauer formalism,
\cite{Landauer1,Landauer2} where the ballistic conductance is given by
$G=G_{0}T(E)$, where $G_{0}$ is the quantum conductance
$G_{0}=2e^2/h$, and $T\left(E\right)$ is the total transmission at the
energy $E$.

$T(E)$ is obtained as
\begin{equation}
T(E) = \sum_{k_\parallel} \sum_{i} \sum_{j}
T_{i,j}\left(k_{\parallel},E\right)
\end{equation}
where $T_{i,j}\left(k_{\parallel},E\right)$ is the probability that an
electron with energy $E$ and transversal momentum $k_{\parallel}$
incoming from the $i-$th Bloch state is transmitted to the outgoing
$j-$th state of the other electrode. Sums runs on both spins.  The
first sum is performed over $k_{\parallel}$ belonging to the
two-dimensional Brillouin zone (2D-BZ) of the supercell.

\begin{figure} [tbp]
\vspace{0.4cm}
\begin{center}
\includegraphics[width=8cm]{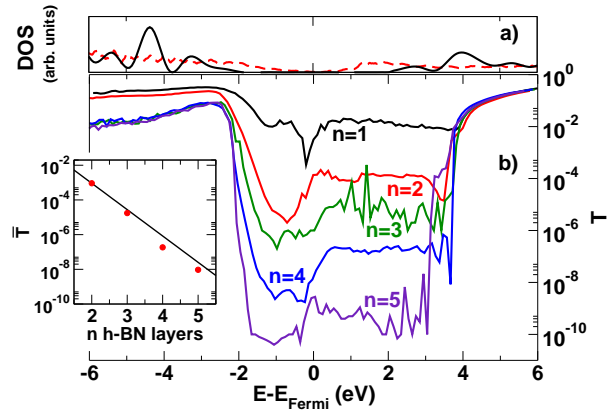}
\end{center}
\caption{Top: Density of states of bulk h-BN (solid black line) and
  graphite (dashed red line). Bottom: Tunneling probability as a
  function of the incident energy for a single h-BN barrier of $n$
  atomic layers between graphite leads}
\label{BN_sb}
\end{figure}

\begin{figure} [tbp]
\vspace{0.4cm}
\begin{center}
\includegraphics[width=8cm]{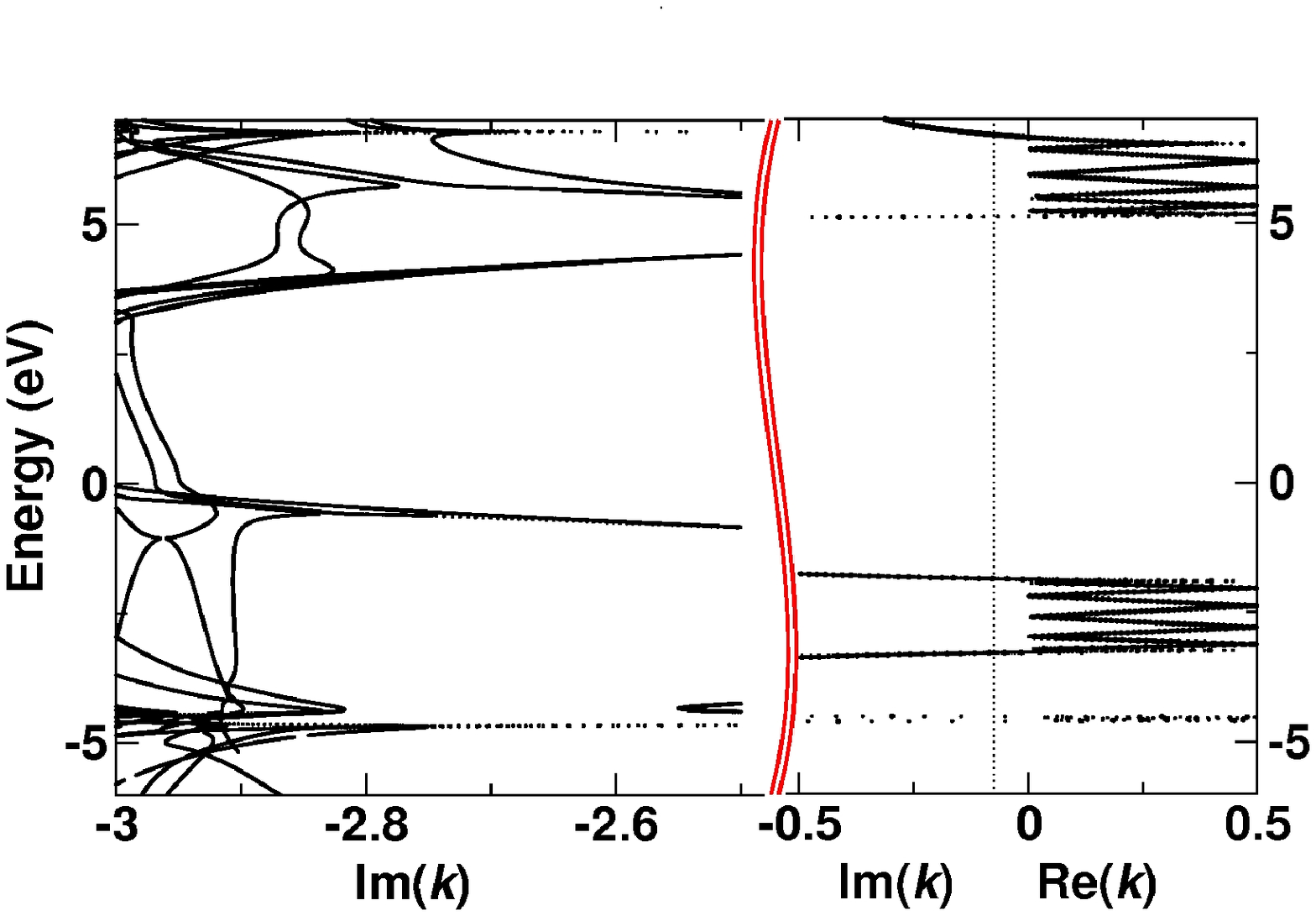}
\end{center}
\caption{Complex band structure for h-BN bulk.}
\label{CB}
\end{figure}

The transmission probability of a single h-BN barrier is shown in
Fig.~\ref{BN_sb} for a different number of layers.  For a given number
$n$ of single h-BN layers, the transmission probability is almost
independent of energy, in an energy range of about 3~eV starting from
the Fermi level. This behavior can be explained by the structure of
the complex bands in bulk h-BN shown in Fig.~\ref{CB}: in the energy
interval considered, in the symmetry point K on the plane parallel to
the layer interface (where the graphite states with real momentum
reside), h-BN has a dispersion relationship according to which the
imaginary wave vector is almost independent of energy, and therefore
$T$ as well.  Of course $T$ decays exponentially with increasing $n$,
as can be verified by plotting the average value $\left(\bar T\right)$
in the energy range between the Fermi energy value and 3.4 eV above as
a function of the number of layers, as in the inset of
Fig.~\ref{BN_sb}. The fitting exponential is

\begin{equation}
\bar T \approx e^{-3.5 n}
\end{equation}

and five h-BN layers strongly suppress the transmission.  The dip in
correspondence of $E-E_{\rm Fermi} = 0$~eV is due to the null density
of states in graphite leads.  In addition, we also note that for
$E-E_{\rm Fermi} < -3$~eV, the transmission probability is suppressed
even in a classically allowed region, probably because of a mismatch
between propagating states in the h-BN barrier and in graphite.

\begin{figure} [tbp]
\vspace{0.4cm}
\begin{center}
\includegraphics[width=8cm]{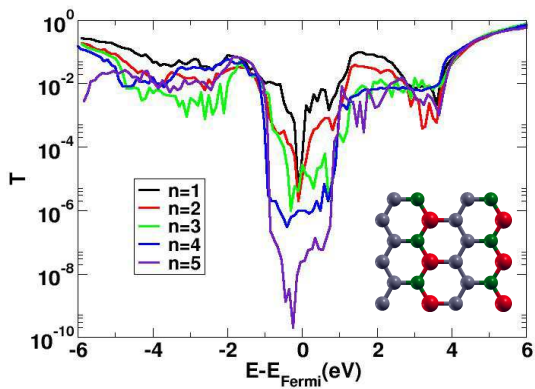}
\end{center}
\caption{Transmission probability as a function of energy for a single
  BC$_{2}$N barrier of $n$ atomic layers between graphite leads. The
  inset shows the atomic structure of the BC$_{2}$N)$_{n}$ isomer
  studied.}
\label{BC2N_sb}
\end{figure}

In Fig.~\ref{BC2N_sb}, the transmission probability as a function of
energy is shown for a single barrier of BC$_2$N of $n$ atomic layers
between graphite leads. The atomic structure is shown in the inset. A
qualitatively similar behavior as h-BN is observed, with an
exponential decay of the transmission probability in a narrow range
around Fermi energy. BC$_2$N has a gap of 1.6 eV.

\begin{figure} [tbp]
\vspace{0.4cm}
\begin{center}
\includegraphics[width=8cm]{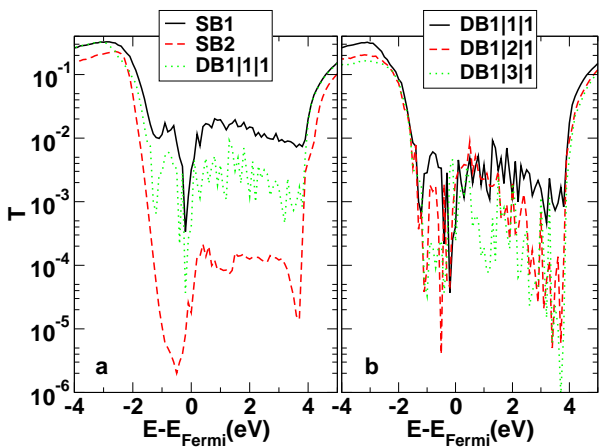}
\end{center}
\caption{Transmission probability as a function of the energy (eV) for
  different systems; {\bf a:} SB1 (solid black line), SB2 (dashed red
  line), DB1$|$1$|$1 (dotted green line). {\bf b:} DB1$|$1$|$1 (solid black
  line), DB1$|$2$|$1 (dashed red line), DB1$|$3$|$1 (dotted green line)}
\label{BN_db}
\end{figure}

Fig.~\ref{BN_db} shows the transmission probability as a function of
energy for some single and double barrier structures.  All these
systems are symmetrical, with two identical barriers separated by one,
two or three graphene layers.  For clarity, hereinafter, we have
indicated the single barrier systems with the acronym SB followed by
the number of h-BN layer (e.g. in
Fig.~\ref{geometry}a, SB5). In the same way, the double barrier systems
are related with DB and three numbers, namely the number of h-BN
layers on the left, the number of graphene layers in the central
region, and the number of h-BN layers on the right (e.g. in
Fig.~\ref{geometry}c, DB2$|$1$|$2).  By comparing the transmission of the
single h-BN barrier (SB1) with that of the double h-BN barrier
(DB1$|$1$|$1) (Fig.~\ref{BN_db}a), the noteworthy aspect is that the
insertion of a single graphene sheet has a noticeable effect on
transport even in the direction perpendicular to the plane
(Fig.~\ref{BN_db}).  While the insertion of even a single BN (SB1)
sheet strongly suppresses the transmittance (particularly in the zone
near the Fermi energy), the presence of a triple layer (DB1$|$1$|$1)
yields almost the same behavior as the single-BN sheet . In addition,
DB1$|$1$|$1 has a much higher transmittance than SB2, despite the
identical number of h-BN layers.  In Fig.~\ref{BN_db}b, we report the
transmittance of double-barrier systems in which two h-BN layers are
separated by one, two or three graphene layers.

\begin{figure} [tbp]
\vspace{0.4cm}
\begin{center}
\includegraphics[width=8cm]{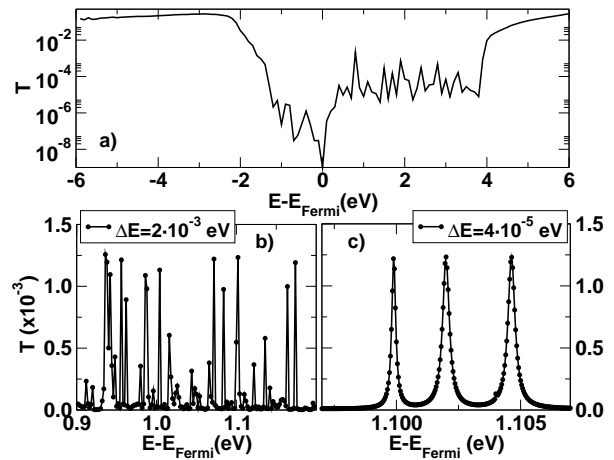}
\end{center}
\caption{a) Transmission probability as a function of the energy (eV)
  for DB2$|$1$|$2; b) and c) Transmission probability as a function of
  energy for DB2$|$1$|$2 system with step sampling $2 \times 10^{-3}$ and
  $4 \times 10^{-5}$.}
\label{212}
\end{figure}

\begin{figure} [tbp]
\vspace{0.4cm}
\begin{center}
\includegraphics[width=8cm]{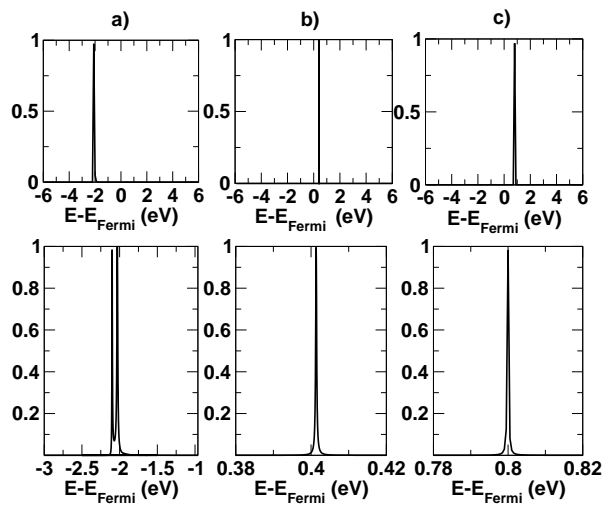}
\end{center}
\caption{Transmission probability of the double barrier system DB2$|$1$|$2
  as a function of energy for selected $k_{\parallel}$: a)
  (0.28000;0.28867); b) (0.31000;0.32909); c) (0.28000;0.33486); in
  unit $\frac{2\pi}{a}$ with a=2.503 \AA }
\label{res212}
\end{figure}

It is also important to notice that no resonant tunneling occurs in
the double barrier structures, in the whole energy range explored, as
we have verified with a very dense energy sampling of transmission
probability, down to steps of $10^{-5}$~eV, as shown in Fig.~\ref{212}
referred to (DB2$|$1$|$2) case.  The reason is due to the fact that the
energy dispersion relation in graphene cannot be written as the sum of
a longitudinal and a transversal component. Therefore, for each
transversal wave vector $k_{\parallel}$, one observes resonances in
$T(E)$ with unity transmission probability peaks, as shown in
Fig.~\ref{res212}, at different energies. When the sum over
$k_{\parallel}$ resonance do not occur at the same energy and are
therefore washed out.

\section{Conclusion}
In this paper we have investigated the tunneling probability of
vertical heterostructures consisting of single and double layers of
h-BN and BC$_2$N separated by graphene sheets and connected to
graphite leads. These structures are interesting from the point of
view of graphene electronics, also because few devices have been
proposed in which transport occurs in the direction perpendicular to
graphene planes~\cite{Britnell,Barristor,Fiori}. As expected, the
transmission probability is exponentially dependent on the number of
layers, and is already strongly suppressed by a single monolayer of
h-BN or BC$_2$N. In addition, due to the energy dispersion
relationship in the energy gap, for which the imaginary wave vector is
almost constant as a function of energy, the transmission probability
has only small dependence on energy.  We have also shown that resonant
tunneling is completely suppressed by the peculiar energy dispersion
relation of graphene, that cannot be decomposed in the sum of a
longitudinal and a transversal component.

\section{Acknowledgment}
This work was supported in part by the EC 7FP through the Project
GRADE (Contract 317839).

\bibliography{PRL}

\end{document}